\documentstyle[multicol,aps,prb,epsf]{revtex}

\newcommand{\Vec}[1]{\mbox{\boldmath$#1$}}

\begin{document}

\draft

\title{Duality and integer
quantum Hall effect in isotropic 3D crystals}
\author{M. Koshino, H. Aoki}
\address{Department of Physics, University of Tokyo, Hongo, Tokyo
113-0033, Japan}

\date{\today}

\maketitle

\begin{abstract}
We show here a series of energy gaps as in Hofstadter's butterfly, which have 
been shown to exist by Koshino et al [Phys. Rev. Lett. {\bf 86}, 
1062 (2001)] for anisotropic 
three-dimensional (3D) periodic systems in magnetic fields $\Vec{B}$, 
also arise in the isotropic case 
unless $\Vec{B}$ points in high-symmetry directions.  
Accompanying integer quantum Hall conductivities 
$(\sigma_{xy}, \sigma_{yz}, \sigma_{zx})$ can, surprisingly, 
take values $\propto (1,0,0), (0,1,0), (0,0,1)$ even for a fixed 
direction of $\Vec{B}$ unlike in the anisotropic case.  
We can intuitively explain the high-magnetic field spectra and the 3D QHE 
in terms of quantum mechanical hopping by introducing 
a ``duality'', which connects the 3D system in 
a strong $\Vec{B}$ with another problem in a weak 
magnetic field $(\propto 1/B)$.

\end{abstract}


\begin{multicols}{2}
\narrowtext
\section{Introduction}
Appearance of energy gaps in three-dimensional(3D) systems is 
a hallmark of quantum mechanics in crystals.  
There, a periodicity, be it atomic or density-wave formation 
from some mechanism, gives rise to a Bragg's reflection 
accompanied by energy gaps as Bloch's theorem dictates.  
Apart from this, there is few occurrences of energy gaps in 
3D (unless of course non-perturbative, many-body 
states such as the BCS state are involved).  

In two-dimensional(2D) systems, by contrast, we do have a 
remarkable realization of gaps when 
a magnetic field is applied.  For a free 2D electron gas the 
spectrum even coalesces into a series of sharp Landau levels.  
When the 2D system has some periodicity (thought of as arising 
from a periodic potential or tight-binding array of atoms), 
application of a magnetic field gives rise 
to a fractal energy gaps, which is called 
``Hofstadter's butterfly''\cite{Hofs}. 
One way to realize why the butterfly appears is that 
the two quantizations, one being Bloch's band structure 
and the other the Landau quantization, 
interfere with each other.  The interference should occur 
when the band gap, $\hbar^2/(m a^2)$
($m$: electron mass, $a$: lattice constant), and 
the cyclotron energy, $\hbar \omega_c$, are similar, which indeed amounts to 
the condition for the appearance of 
the butterfly, $\phi \sim \phi_0$, where $\phi$ is the magnetic 
flux penetrating a unit cell and $\phi_0=h/e$ the flux quantum.  

Once we have an energy gap in a 2D system it is known that 
the integer quantum Hall effect (IQHE) should occur.
Thouless {\it et al}\cite{Thou} 
in fact have derived a general formula for the
quantized Hall conductance in 2D periodic systems 
without any assumption on the periodicity,
and shown that the Hall conductance can be written in terms of 
some topological invariants.  

A natural question is whether we can extend such arguments to 
3D systems.  It has been shown that, {\it if} 
an energy gap exists in 3D, it 
should accompany an IQHE where each component of the Hall tensor 
$\sigma_{ij}$ is quantized\cite{Halp,Mont,Kohm,Goryo}.  
A big question, then, is whether and how gaps can appear in 3D.  
We have previously shown that the butterfly-like
spectrum indeed appears in {\it anisotropic} 3D (quasi-1D) lattice and 
have derived the quantized Hall tensor for each gap.\cite{Kosh}  
For the anisotropic case we can intuitively understand these in 
terms of a mapping between the 3D case and a 2D case.  
For the isotropic case, however, 
it seems difficult to have the energy gaps 
since the magnetic subbands are likely to be overlapped
by the large dispersion along the magnetic field.
Several authors \cite{Hase,Kuns} have computed 
the energy spectra of the isotropic 3D tight-binding system
under magnetic fields pointing to high-symmetry 
crystallographic directions, 
and Kunszt\cite{Kuns} concluded that there is in general no energy gaps
although the possibility of gaps cannot be excluded.
So the existence of gaps and IQHE remains an open question.

In this paper, we show that we do have energy gaps 
in {\it isotropic} 3D lattices rather universally in that 
all we have to have is the sufficiently large 
magnetic field $\Vec{B}$ pointing to 
general directions (i.e., off the high-symmetry axes).  
We first show this by calculating energy spectra for 
various directions of $\Vec{B}$.  
We have also calculated the Hall tensors, 
$\sigma_{xy}, \sigma_{yz}, \sigma_{zx}$, when the Fermi energy
is in each energy gap.  They turn out to have values, 
e.g. $(1,0,0),(0,1,0),(0,0,1)$ in 
an appropriate unit, 
even for a fixed direction of $\Vec{B}$, which 
is surprising and unlike in the anisotropic case.

We then intuitively explain such structures in the spectra 
by introducing a ``duality'', which connects the 3D system in 
a strong $\Vec{B}$ with another problem in a weak 
magnetic field $(\sim 1/B)$. This enable us to regard the 
band structure and the quantization of the 3D Hall conductivity
as graphically arising from quantum mechanical hopping 
between ``semiclassical'' orbits in the weak-$B$ case.

\section{Formulation of 3D Bloch electrons in magnetic fields}
We take a non-interacting tight-binding model 
in a uniform magnetic field $\Vec{B}=(B_x,B_y,B_z)$ 
pointing to an arbitrary direction.
Schr\"{o}dinger's equation is 
$-\sum_j t_{ij} e^{i\theta_{ij}}\psi_j = E\psi_i$,
where $\psi_i$ is the wave function at site $i$,
the summation is over nearest-neighbor sites, and 
$\theta_{ij} = \frac{e}{\hbar}\int_{j}^{i}\Vec{A}\cdot {\rm d}\Vec{l}$ 
is the Peierls phase with \Vec{A} the vector potential, 
$\nabla\times\Vec{A} = \Vec{B}$. 
We consider a simple-cubic lattice with a lattice constant $a$.  
Following Kunszt and Zee,\cite{Kuns} we take 
a gauge $\Vec{A}=(0, B_z x - (B_zB_x/B_y)(y-a/2), B_x y - B_y x)$, 
where $z$ is cyclic.  We can then write 
$\psi_{lmn} = e^{i \lambda_z n} F_{l,m},$ 
where $l,m,n$ are site indices along $x,y,z$, respectively.
Schr\"{o}dinger's equation is then reduced to a two-dimensional 
tight-binding model,
\begin{eqnarray}
  &-&t_x F_{l-1,m} \,-\, t_x F_{l+1,m} 
  -t_y e^{2\pi i \phi_z [l-(\phi_x/\phi_y)(m-1)]}F_{l,m-1} \nonumber \\ 
  &-&t_y e^{-2\pi i \phi_z [l-(\phi_x/\phi_y)m]}F_{l,m+1}\nonumber \\
  &-& 2t_z \cos[2 \pi (\phi_x l - \phi_y m) + \lambda_z] F_{l,m} = E F_{l,m}
  \label{Harp3D},
\end{eqnarray}
where $t_i (i=x,y,z)$ is the transfer energy and 
$\phi_i = B_i a^2/\phi_0$ with $\phi_0 \equiv h/e$ 
is respective number of flux quanta
penetrating the facet of the unit cell normal to $\hat{\Vec{e}}_i$.  
Here we consider rational fluxes, 
\begin{equation}
(\phi_x,\phi_y,\phi_z) = \Phi \times (n_x,n_y,n_z), \,\,\, \Phi = P/Q  
\end{equation}
with $P,Q$: integers and $n_i$'s are mutually prime integers.  
By introducing $j = n_y l - n_x m$, 
we can eliminate $m$ from the phase factors (exponentials and 
the argument of cosine in Eq. (\ref{Harp3D})), so we can write 
$F_{j,m} = e^{i \lambda_y m}G_{j}$, where $G$ is determined 
from a one-dimensional equation, 
\begin{eqnarray}
  &-&t_x G_{j+n_y} \,-\, t_x G_{j-n_y} \nonumber \\ 
  &-&t_y e^{i[2\pi \Phi \frac{n_z}{n_y}(j+n_x)-\lambda_y]}
  G_{j+n_x} 
  -t_y e^{-i(2\pi \Phi \frac{n_z}{n_y} j-\lambda_y)}
  G_{j-n_x} \nonumber \\
   &-& 2t_z \cos( -2 \pi \Phi j - \lambda_z ) G_{j} = E G_{j}
  \label{HarpReduced}.
\end{eqnarray}
Since the phase factors have a common periodicity $n_y Q$,
we can apply the Bloch-Floquet theorem to have 
$G_{j+n_y Q} = e^{i\lambda_x n_y Q}G_{j}$, and
the Hamiltonian is reduced to a $n_y Q\times n_y Q$ matrix.

\section{Energy spectra and Hall conductivities}
\begin{figure}
\begin{center}
  \leavevmode\epsfxsize=150mm \epsfbox{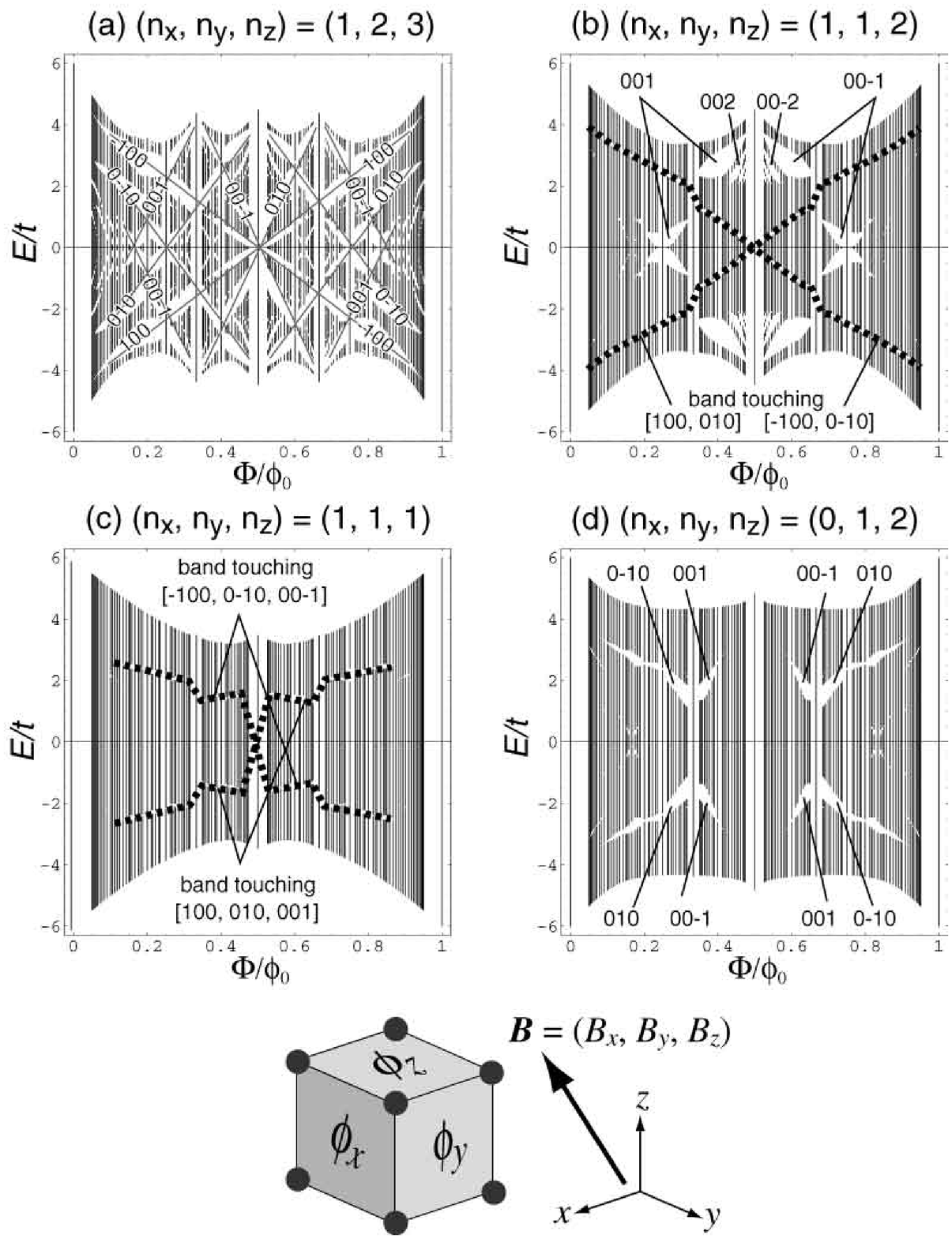}
\end{center}
\caption{Energy spectra for the isotropic simple-cubic lattice 
in the magnetic flux 
$(\phi_x,\phi_y,\phi_z) = \Phi \times (n_x,n_y,n_z)$ 
(as depicted in the inset).  Each spectrum is
plotted against $\Phi$
with the direction of $\Vec{B}$ fixed at 
(a)$(n_x,n_y,n_z)=(1,2,3)$, (b)(1,1,2),
(c)(1,1,1), (d)(0,1,2).  Triple integers represent 
the Hall integers, $(\sigma_{xy}, \sigma_{yz}, \sigma_{zx})$ 
in units of $(-e^2/h a)$.  
Dashed lines in (b),(c) delineate the zero-gap.  
} 
\label{fig_btfl}
\end{figure}

Our aim is to search systematically for gapful spectra 
for the isotropic case with $t_x = t_y = t_z \equiv t$ and to
determine the Hall conductivities for each energy gap.  
We have numerically solved Eq.(\ref{HarpReduced}) 
and obtained the energy spectra versus $\Phi $.  
Fig. \ref{fig_btfl} shows the results for typical field directions:
$(n_x,n_y,n_z)=(1,2,3), (1,1,2), (1,1,1)$, and $(0,1,2)$,
which differ in symmetry.
In each case we took 16 points for each of the Bloch wavenumbers 
$\lambda_x, \lambda_y, \lambda_z$. 

We can immediately see that a series of gaps appear 
for $(n_x,n_y,n_z)=(1,2,3)$, while otherwise we have at most 
solitary gaps.  From such results we have found 
that the spectrum has a series of gaps 
{\it when no two $n_i$'s coincide and $n_i \neq 0$}. 
So we can say that the gapful spectrum can be expected
when $\Vec{B}$ points to general directions 
(i.e., off high-symmetry crystallographic axes).  
The gaps shrinks when $B$ is too small ($Ba^2/\phi_0 \ll 1$) obviously.

Now we proceed to the quantum Hall effect.  As mentioned, 
$E_F$ in an energy gap dictates\cite{Halp,Mont,Kohm} quantized 
Hall conductivities, 
\begin{equation}
(\sigma_{yz},\sigma_{zx},\sigma_{xy})=-\frac{e^2}{ha}(m_x,
m_y,m_z), 
\label{3DQHE}
\end{equation}
where $m_i$'s are integers 
satisfying a Diophantine's equation,
\begin{equation}
 \nu_B = s + m_x \phi_x + m_y \phi_y + m_z \phi_z.
\label{Dio3D}
\end{equation}
Here $s$ is an integer, 
$\nu_B$ is the filling of the tight-binding band, and 
$m_i$'s assigned to each gap are topological invariants, i.e.,
they never change when external parameters (magnetic field, 
transfer energies, etc) are changed as long as the gap remains.  
So Eq.(\ref{Dio3D}) tells us that, if we consider $\nu_B$ as a function 
of $(\phi_x,\phi_y,\phi_z)$ in a 4D parameter 
space $(\nu_B,\phi_x,\phi_y,\phi_z)$, 
the $\nu_B$ plotted along a gap assigned with $(m_x,m_y,m_z)$ will 
form a `3D plane' having a gradient $(m_x,m_y,m_z)$ in the 4D space.

In fact we can readily translate the $E$-$\Phi $ diagram in Fig. \ref{fig_btfl}
to the $\nu_B$-$\Phi $ diagram by counting the number of subbands below 
each value of $E$.   
There, the gradient $\partial \nu_B / \partial \Phi$
for each gap gives $ m_x n_x + m_y n_y + m_z n_z$.  
We can then scan the direction of $\Vec{B}$ to determine
all of $m_x,m_y,m_z$ by keeping track of each gap in question, 
which is exactly how we have obtained the Hall
integers here.
The result is shown as triple integers in Fig. \ref{fig_btfl}. 
We can see that the largest series of gaps have 
$(m_x,m_y,m_z)=(\pm 1,0,0),(0,\pm 1,0)$ or $(0,0,\pm 1)$,
where only one of the Hall components is nonzero.  
This is strikingly new in that one and the same 
butterfly contains all the cases with $m_x\neq 0, m_y\neq 0, m_z\neq 0$, 
respectively, for a fixed direction 
of $\Vec{B}$.  This is in sharp contrast with 
the butterfly in the anisotropic case\cite{Kosh} 
where one component $\sigma_{yz}$ is identically zero 
with $x$ being the most conductive direction.  
Hence the versatile behavior of $(m_x,m_y,m_z)$ may be 
regarded as a hallmark of the isotropic 3D QHE.

When the above condition is not satisfied (i.e., 
when two $n_i$'s coincide 
as in $(1,1,2)$ and $(1,1,1)$ or at least one $n_i= 0$ as in $(0,1,2)$),
only solitary gaps or zero gaps 
(marked with dashed lines) appear.  
Hasegawa has shown for the case of $(1,1,1)$ 
that two bands touch where the energy dispersion 
is similar to those of a Dirac particle. 
We have similar band touching in Fig.\ref{fig_btfl}(b)$(1,1,2)$ 
as well, for which a detailed calculation shows that 
the dispersion around the band touching is indeed 
linear in $\lambda_x, \lambda_y, \lambda_z$.  

The reason for the band touching can be
accounted for as follows.  For $(n_x,n_y,n_z) =
(1,1,2)$ (b), the gap for $(m_x,m_y,m_z) = (m,0,0)$ and that for 
$(0,m,0)$ appear along the same line in the $\nu_B$-$\Phi $ diagram, 
since the problem is symmetric 
against $x\leftrightarrow y$.  If there were no band touching, 
the Hall integers for $(m,0,0)$ and 
$(0,m,0)$ would become indefinite, 
which contradicts with the topological argument 
that we can always
determine $m_i$'s uniquely when the gap is nonzero.  So we can conclude that
the band touching always occurs when two $n_i$'s coincide.  This
explains why the series of gaps seen in case (a) disappear 
as we make $n_x=n_y$ in (b) and then $n_x=n_y=n_z$ in (c).  
We still have to 
explain why the butterfly does not appear in case (d) 
with one of $n_i$'s being zero.
We address this question from a different viewpoint in the next section.

\newpage 

\newpage
\section{Duality for $B \leftrightarrow 1/B$}
Electrons in magnetic fields are usually described 
in $\Vec{k}$-space.  A textbook example is the
semiclassical picture, where an electron is treated as a classical
particle driven in $\Vec{k}$-space by a weak magnetic field.  
As $B$ becomes stronger, the semiclassical orbits 
begin to be mixed and the picture breaks down.
Actually this quantum mixing plays a critical role 
in opening the gaps and in quantization of the Hall conductivity
as shown in the following.

We propose here that there exists a duality that relates a semiclassical
picture in $\Vec{k}$-space (weak-$B$ case) to another semiclassical picture in
the opposite limit of strong magnetic field, in which the {\it guiding
center} of the cyclotron motion drifts along equipotential contours
in the real space.
Using a quantum mechanical mapping from the weak-$B$ case to the
strong-$B$ case (where $B$ translates into $1/B$), we are able to
discuss the mixing of semiclassical orbits in our problem 
(3D lattice in $\Vec{B}$) in a clear cut.

Let us derive the duality in 2D and then extend it to 3D.
We consider a 2D Bloch electron with the dispersion
$\varepsilon(k_x,k_y)$ in a magnetic field $B$ perpendicular to the plane.
If we define the dynamical wavevector 
$\Vec{K} \equiv \Vec{k} + (e/\hbar)\Vec{A}$,
the Hamiltonian in $k$-space is written as
\begin{eqnarray}
 && H = \varepsilon(K_x,K_y),\nonumber \\
 &&[K_x,K_y] = -ieB/\hbar.
\label{KxKy}
\end{eqnarray}
This is an exact quantum equation except that we have 
neglected the inter-band mixing due to the magnetic field,
which is allowed when the periodic potential is strong enough
(or the magnetic field is weak enough).
The equation reduces to the semiclassical one 
in the limit $eB/\hbar \rightarrow 0$.

If we consider another 2D system in a magnetic field $B$ 
with a periodic potential $V(\Vec{r})$ with $\Vec{r}\equiv (x,y)$, 
the Hamiltonian in real space is written as
\begin{eqnarray}
 H &=& \frac{1}{2m}(\Vec{p}+e\Vec{A})^2 + V(\Vec{r})\nonumber\\
 &=& \frac{1}{2m}\Vec{\Pi}^2 
+ V\left(\Vec{R}-\frac{1}{eB}\hat{\Vec{e}}_z\times \Vec{\Pi}
\right),
\label{XY-1}
\end{eqnarray}
where $\Vec{\Pi} \equiv \Vec{p} + e\Vec{A}$ and
$\Vec{R}=\Vec{r}-\Vec{\xi}$ is the cyclotron-motion guiding center 
with the relative coordinate given as 
$\Vec{\xi} = -(1/eB)\hat{\Vec{e}}_z\times \Vec{\Pi}$.

When the magnetic field is strong enough 
(or the potential $V$ is weak enough) 
so that (i) the magnetic length $l = \sqrt{\hbar/(eB)}$
is much shorter than the length scale over which 
$V$ varies, and that (ii) the different Landau levels are not mixed, 
we can consider $\Vec{\Pi}$ to be frozen.   More precisely, 
the kinetic energy (the first term in Eq.(\ref{XY-1}))
reduces to a constant from (ii), while the second term can be 
approximated as $V(\Vec{R})$ from (i).  
The Hamiltonian then becomes, up to a constant, 
\begin{eqnarray}
 && H \approx V(X, Y),\nonumber \\
 &&[X,Y] = i\hbar/(eB).
\label{XY-2}
\end{eqnarray}
We can see that Eqs. (\ref{KxKy}) and (\ref{XY-2})
are identical if we translate as
\begin{equation}
 \Vec{K} \leftrightarrow \Vec{X}, \,\,\,\,\,\,
 \varepsilon \leftrightarrow V, \,\,\,\,\,\,
 \frac{eB}{\hbar} \leftrightarrow \frac{\hbar}{eB}.
\label{duality}
\end{equation}
So we can see that there exists a duality between
the 2D system with a strong periodic potential in a weak $B$ 
(one band tight-binding model) and
the 2D system with a weak periodic potential in a strong field
(one Landau level approximation).  
For the Landau level in the latter case we expect that 
the duality holds best in the lowest Landau level, since 
the wavefunction is most localized in that level 
to make Eq. (\ref{XY-2}) valid.

The duality introduced here provides a 
physical basis for Hofstadter's observation\cite{Hofs} on 
the mapping between the tight-binding and 
weak-potential systems\cite{Rauh} for the square lattice, and also
generally applicable to any 2D lattices, or, remarkably, 3D cases as well 
as we shall see below.  
Another duality is found by Ishikawa {\it et al}\cite{Ishi}, which 
is distinct from ours in that
one Landau level is mapped to one Landau level in the latter 
while one Landau level is mapped to one Bloch band in the 
present case.

Now we apply the duality to 3D systems, which 
enables us to have a fresh look at the 3D QHE.  
We take the Bloch electron with the dispersion 
$\varepsilon(k_x,k_y,k_z)$ in a magnetic field
$\Vec{B} = (B_x,B_y,B_z)$. Let us again denote 
the direction of $\Vec{B}$ as $\Vec{n}\equiv (n_x,n_y,n_z)$ 
whose components will 
be assigned a set of integers.
We set a new frame $(x',y',z')$ 
in which  the magnetic field point to $(0,0,1)$.
If we take the vector potential $\Vec{A}$ with $A_{z'} = 0$,
we have
\begin{equation}
 H = \varepsilon(K_{x'},K_{y'},k_{z'}),
\end{equation}
where $K_i = k_i + (e/\hbar)A_i$ $(i=x',y')$ with
$[K_{x'},K_{y'}] = -ieB/\hbar$ while $k_{z'}$ remains a quantum number.
We can thus apply the above duality to this system, which implies that 
the 3D system can be mapped to a 
system in a fictitious 2D space $(K_{x'},K_{y'})$ 
with a potential $\varepsilon(K_{x'},K_{y'},k_{z'})$
in a fictitious magnetic field $B^*$ with $\hbar/(eB^*) = eB/\hbar$.
The mapping is valid for a large $B^*$, i.e., a small $B$.
The fictitious potential is a 
cross section of the 3D dispersion incised at $k_{z'}$ by a 
plane perpendicular to $\Vec{B}$. 
As we shift $k_{z'}$, the 2D potential changes accordingly, and 
the spectrum is determined by taking all the possible values of $k_{z'}$.

We take a cubic dispersion,
\begin{equation}
 \varepsilon(k_x,k_y,k_z) 
= - t_x \cos k_x a - t_y \cos k_y a - t_z \cos k_z a, 
\end{equation}
which is written in the new frame as,
\begin{eqnarray}
 \varepsilon(k_{x'},k_{y'},k_{z'}) =
&-& t_x \cos[\Vec{G}_x \cdot \Vec{k}_{\perp} + (n_x/n)k_{z'} a] \nonumber\\
&-& t_y \cos[\Vec{G}_y \cdot \Vec{k}_{\perp} + (n_y/n)k_{z'} a] \nonumber\\
&-& t_z \cos[\Vec{G}_z \cdot \Vec{k}_{\perp} + (n_z/n)k_{z'} a],
\label{vareps}
\end{eqnarray}
where $\Vec{k}_{\perp} \equiv (k_{x'},k_{y'})$, 
$n=\sqrt{n_x^2+n_y^2+n_z^2}$ and
\begin{eqnarray}
\Vec{G}_x &=& \tilde{a} (n_z n_x/n, -n_y),\nonumber\\
\Vec{G}_y &=& \tilde{a} (n_y n_z/n, n_x),\nonumber\\
\Vec{G}_z &=& \tilde{a} (-(n_x^2+n_y^2)/n, 0),
\end{eqnarray}
with $\tilde{a} \equiv a/\sqrt{n_x^2+n_y^2}$.
Now we can see that
our 3D problem is reduced to a fictitious 2D system having three periods 
[while the Hofstadter problem (2D Bloch system in $B$)
is reduced to a 1D system having two periods].  
The Hamiltonian for the fictitious 2D system is 
\begin{equation}
 H = \frac{1}{2m}(\Vec{p}^* + e \Vec{A}^*)^2 
+ \varepsilon(K_{x'},K_{y'},k_{z'}),
\end{equation}
where $\Vec{p}^* = (\partial/\partial K_{x'}, \partial/\partial K_{y'})$
and $\Vec{A}^*$ is a vector potential for the
fictitious magnetic field $(0,0,B^*)$.
Since we assume a large $B^*$,
Schr\"{o}dinger's equation can be written within the basis 
for the lowest Landau level, 
which reads, after a certain phase transformation,
\begin{eqnarray}
  &-&\tilde{t}_x \Psi_{j+n_y} \,-\, \tilde{t}_x \Psi_{j-n_y} }
  -\tilde{t}_y e^{i [2\pi \Phi \frac{n_z}{n_y}(j+\frac{n_x}{2})-\lambda_y]}
  \Psi_{j+n_x \nonumber \\ 
  &-& \tilde{t}_y e^{-i [2\pi \Phi \frac{n_z}{n_y}(j-\frac{n_x}{2})-\lambda_y]}
  \Psi_{j-n_x} \nonumber \\
   &-& 2\tilde{t}_z \cos( -2 \pi \Phi j - \lambda_z) \Psi_j = E \Psi_j.
\label{HarpReduced2}
\end{eqnarray}
Here 
the hopping parameters are
\begin{equation}
\tilde{t}_i = t_i e^{-G_i^2 l^{*2}/4}, 
\end{equation}
$l^* \equiv \sqrt{\hbar/(eB^*)}$ is the fictitious magnetic length, and 
$\lambda_y, \lambda_z$ are given by
\begin{eqnarray}
 \lambda_y &=& -\Phi \frac{n_z}{n_y} \alpha 
  + \frac{n_x^2+n_y^2}{n n_y} k_{z'} a \nonumber\\
 \lambda_z &=& -\Phi \alpha + \frac{n_z}{n} k_{z'} a,
\end{eqnarray}
where $\alpha$ labels the wavefunction
in the lowest Landau level.
Now the duality implies that 
Eq.(\ref{HarpReduced2}) should reduce to Eq.(\ref{HarpReduced})
in the strong $B^*$ limit (i.e., weak $B$ limit), 
since $l^*$ should become smaller than the potential range $1/G_i$ 
so that $\tilde{t}_i$ tends to $t_i$.

\section{Interpretation of the 3D QHE from the duality}

The mapping via the duality 
enables us to see how the total spectrum comes from 
dispersion relations versus $k_{z'}$, which 
helps intuitively understanding how 
the energy gaps open and how the Hall conductivities are quantized.
Since the mapping is limited to the weak-$B$ case,
we examine the situation where the gaps begin to open 
in the quantum regime approached from the semiclassical one.
We show in Fig.\ref{fig_band} the band structure versus
$k_{z'}$ for the fictitious system (Eq.(\ref{HarpReduced2})), 
along with the fictitious potential
$\varepsilon(K_{x'},K_{y'},k_{z'})$ for fixed values of $k_{z'}$.  
Let us first look at how the energy spectra
can have gaps for general directions of $\Vec{n} [\propto (1,2,3)$ in 
Fig.\ref{fig_band}(a)].  $\varepsilon$ accommodates closed orbits
around its peaks and dips in the ``semiclassical picture'' in 
the language for the weak-$B$ (strong-$B^*$) case, where the area
enclosed by each orbit must be a multiple of $2\pi /l^2$.  So the
different wells have different sets of discrete levels, where each
level moves on the energy axis as $k_{z'}$ is changed, since
$\varepsilon$ changes its form with $k_{z'}$.  When the levels belonging to 
the wells adjacent in $K_{x'}K_{y'}$-space coincide, the states will 
strongly resonate quantum-mechanically and an energy gap will arise.
In Fig.\ref{fig_band}(a), we show the case where the gaps begin to open with 
$(\phi_x,\phi_y,\phi_z)=(1/9)(1,2,3)$, for which the strong mixing of 
orbits is seen to result in significant level repulsions.  
We attach for comparison the result 
for an almost semiclassical case when the 
magnetic field, $(\phi_x,\phi_y,\phi_z) = (1/45)(1,2,3)$, is so small that  
the mixing is weak and the level repulsion is almost 
negligible except for the middle of the total band. 

In terms of the semi-classical orbits, the change in $k_{z'}$ 
causes the orbits to hop to adjacent positions 
at every resonance as shown in Fig.\ref{fig_band}.  
A virtue of this picture is that 
{\it the distance and the direction of the hopping
exactly indicates the Hall integers}, 
as shown in the following,
If we apply an infinitesimal
electric field $\Vec{E}$ to the system, $k_{z'}$ is dragged
adiabatically according to an equation, $\hbar (dk_{z'}/dt) = -e E_{z'}$, 
where $\Vec{E}_{z'}$ is the component parallel to $z'$.  After
$k_{z'}$ is changed by $\delta k_{z'} \equiv 2\pi/(an)$ 
(the period with which the spectrum repeats
itself), every state must come back to the 
equivalent position in the reciprocal unit cell.  
Note here that the cell boundaries (represented in Fig.\ref{fig_band} 
as white lines) also move as $k_{z'}$ is changed, since 
the Brillouin zone boundaries are oblique with $z'\parallel \Vec{B}$.  
The increment in $k$-space, 
$\delta\Vec{K} = (\delta K_{x'},\delta K_{y'},\delta k_{z'})$, 
with which the orbit is shifted over the one period satisfies
\begin{eqnarray}
\delta\Vec{K} &=& \mp \frac{2\pi}{a} \delta\Vec{m},\nonumber\\ 
\delta\Vec{m} &=& \delta m_x \hat{\Vec{e}}_x+ \delta m_y \hat{\Vec{e}}_y + \delta m_z \hat{\Vec{e}}_z,
\end{eqnarray}
where $-$ and $+$ correspond to positive and negative
$E_{z'}$, respectively,
and $\delta m_i$'s are integers assigned to each subband.
The integers satisfy a relation 
$\delta \Vec{m} \cdot \Vec{n} =1$, since $\delta k_{z'} = 2\pi/(an)$. 
We can immediately translate $\delta\Vec{K}$ 
into the motion in the real
space normal to $z'$ using the relationship between 
the relative coordinate $\Vec{\xi}$ and the dynamical wavenumber $\Vec{K}$,
\begin{equation}
\delta\Vec{\xi} = 
-\frac{\hbar}{eB}\hat{\Vec{e}}_{z'} \times \delta\Vec{K}.
\end{equation}
Since the system has no dissipation as long as $E_F$ is in a gap,
$\Vec{E}_{z'}$ should cause no net current along $z'$.  
Therefore the velocity averaged over on period 
driven by $\Vec{E}_{z'}$ is just the ratio 
of $\delta\Vec{\xi}$ and the one period ($=\delta k_{z'}/(e|E_{z'}|/\hbar))$, 
which leads to
$ \Vec{v}
= (n/B)\Vec{E}_{z'}\times \delta\Vec{m}$.
The Hall current due to $\Vec{E}_{z'}$ is calculated as
\begin{equation}
 \Vec{j}
 = -\rho e \Vec{v}
 = \frac{e^2}{ha}\delta\Vec{m}\times\Vec{E}_{z'},
\end{equation}
where $\rho = eB/(ahn)$ is the density of states per subband
and per unit volume.  
From this expression we can write the Hall tensor as
$\hat{\Vec{\sigma}}_{\perp} = -(e^2/ha)\delta\Vec{m}_{\perp}$,
where $\sigma_{ij} \equiv \epsilon_{ijk}\hat{\sigma}_k$,
and $\perp$ represents the component normal to $z'$.
On the other hand, the normal component of the electric field,
$\Vec{E}_{\perp}$ causes a classical drift in the direction
normal to $z'$ with the velocity
$\Vec{v}
= (\Vec{E}_{\perp} \times \Vec{B})/B^2 
= (n/B) \Vec{E}_{\perp}\times \delta\Vec{m}$,
which immediately leads to 
$\hat{\sigma}_{z'} = -(e^2/ha)\delta m_{z'}$.
Combining the two, we finally obtain the quantized Hall conductivity
carried by the corresponding subband, 
\begin{equation}
\hat{\Vec{\sigma}} = -(e^2/ha)\delta\Vec{m}. 
\end{equation}
We have thus derived the quantization of the 3D Hall conductivity
from the $\Vec{k}$-space hopping, as an approach alternative 
to the usual Kubo formula.
The problem of the transport in adiabatically varying potentials 
(so-called Thouless pumping) was first considered by 
Thouless\cite{ThouPump} for one-dimensional case.  
He showed that the 2D QHE in periodic potentials may be understood 
in terms of a fictitious 1D system having 
two periodic potentials that slide adiabatically with each other.
The discussion we have given here to describe the 3D QHE
may be regarded as a two-dimensional version of the Thouless pumping.

We can actually identify the Hall integers $\delta m_i$
for each subband by keeping track of the motion of 
the orbits with $k_{z'}$. Figure 2(a) typically depicts how 
a state in the first band moves by (1,0,0), 
by which we mean the state jumps once 
across a reciprocal cell boundary normal to $k_x$  
(whose intersection is shown as one of the white lines 
in Fig.2) but not across $k_y$ or $k_z$.  
The state in the second band moves by $(-1,1,0)$.
These triple numbers are the very Hall integers, 
$(\delta m_x,\delta m_y,\delta m_z)$, carried by each subband 
(not to be confused with the total Hall integer),
and are in accordance with the small
$\Phi$ region in Fig.\ref{fig_btfl}(a).

We can now comment on the relation with the usual treatment of 3D 
QHE\cite{Halp,Mont,Kohm}.  
By summing up $\delta \Vec{m} \cdot \Vec{n} = 1$ over the
occupied subbands, we have
$\Vec{m} \cdot \Vec{n}=r$, 
where $r$ is the number of occupied subbands 
and $\Vec{m}$ is the summation of $\delta\Vec{m}$ 
over them. Since $r$ is related to the filling of the tight-binding band in
$\nu_B = r\times\rho/(1/a^3)$, 
we obtain 
\begin{equation}
 \nu_B = m_x \phi_x + m_y \phi_y + m_z \phi_z,
\end{equation}
which coincides with the Diophantine equation (\ref{Dio3D}) for $s=0$.  
This means that in our picture above,
only the gaps with $s=0$ are taken into account while the rest are neglected.
We can actually show that
the gap associated with $s\neq 0$ only occurs, 
when the magnetic field is weak ($\Phi \ll 1$), 
as a small gap between very dispersive bands (versus $k_{z'}$), 
so that the gap vanishes in the spectrum.  
Weak field regime is exactly the situation considered here, 
since we are approaching to the quantum regime from the semiclassical one.
The discussion here is for the lower half of
the tight-binding band, while we can make a similar 
argument for the upper half in terms of 
hole orbits, which can be obtained as an electron-hole 
transformation (on the $\nu_B$-$\Phi $ diagram), 
which in turn corresponds to $s = 0 \rightarrow 1$ in the Diophantine
equation.

The mapped picture also gives an intuitive explanation why we have 
so few gaps for the symmetric case Fig.\ref{fig_band}(b),(c),
and zero-component case (d).
When two components in $\Vec{n}$ coincide as in Fig.\ref{fig_band}(b) 
with $\Vec{n} \propto (1,1,2)$, while two
levels cross (for the value of $k_{z'}$ labeled 
as B), a gap does not arise because the
couplings along $k_x$ and $k_y$ occur symmetrically (in a 
zigzag fashion), so that the bands
do not split, which is exactly the band touching discussed in Sec. III.
When the symmetry is even higher 
with $\Vec{n} \propto (1,1,1)$ in Fig.\ref{fig_band}(c), 
couplings along $k_x$, $k_y$ and $k_z$ all become symmetric 
and the band touching occurs at every energy crossing.
If $\Vec{n}$ contains a zero component as in (d), 
two of the plane waves for $\varepsilon$ become
parallel and the wells having the same depth become connected along the
perpendicular direction. This results in a strong mixing between the states
along that trough direction, so that the minibands for each value of $k_{z'}$
become wider.  So, while the energy gap does arise when
the energies of the adjacent troughs coincide, these gaps tend to be 
overlapped in energy by other, wide bands.  

\begin{figure}
\begin{center}
  \leavevmode\epsfxsize=80mm \epsfbox{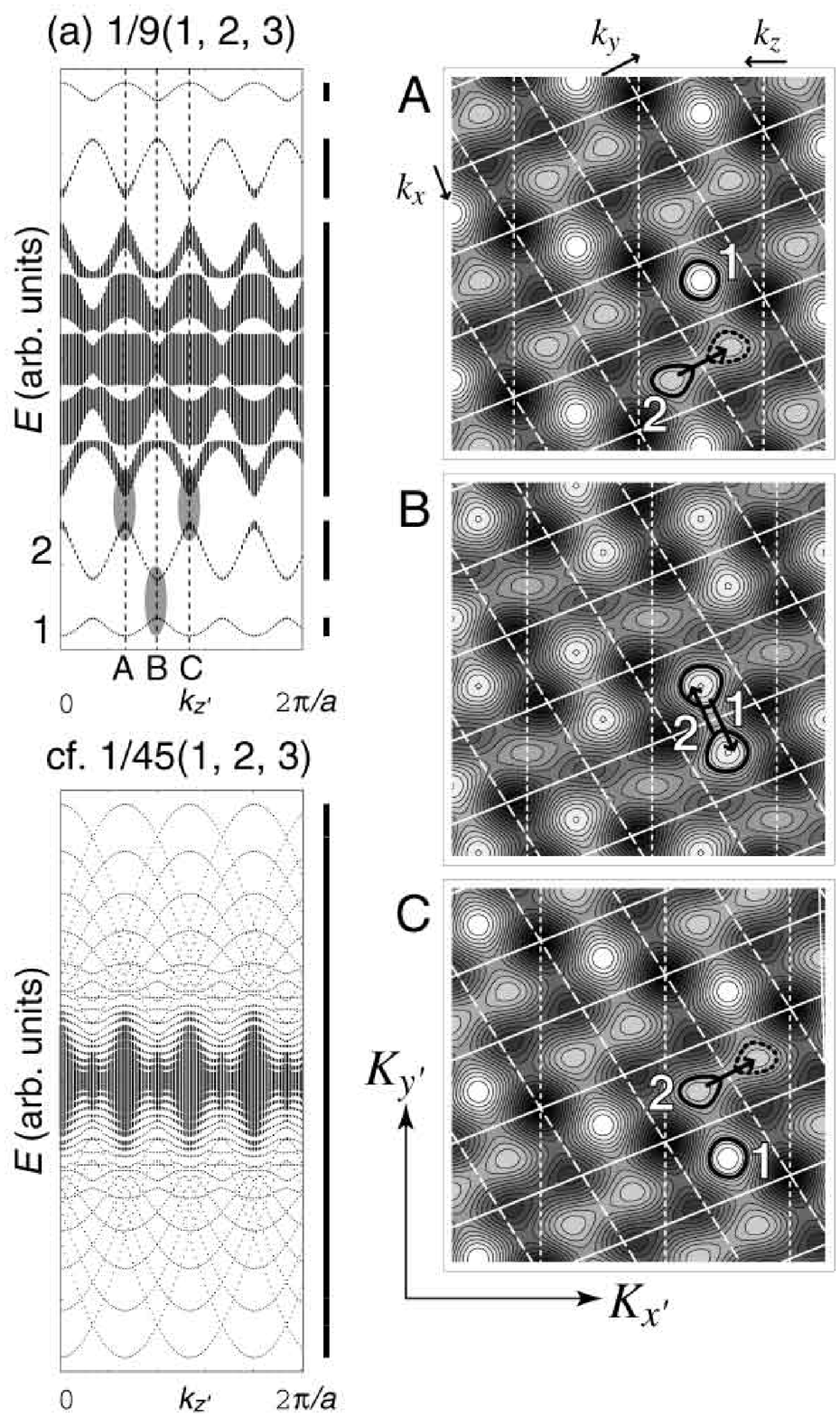}
\end{center}
\end{figure}
\begin{figure}
\begin{center}
  \leavevmode\epsfxsize=80mm \epsfbox{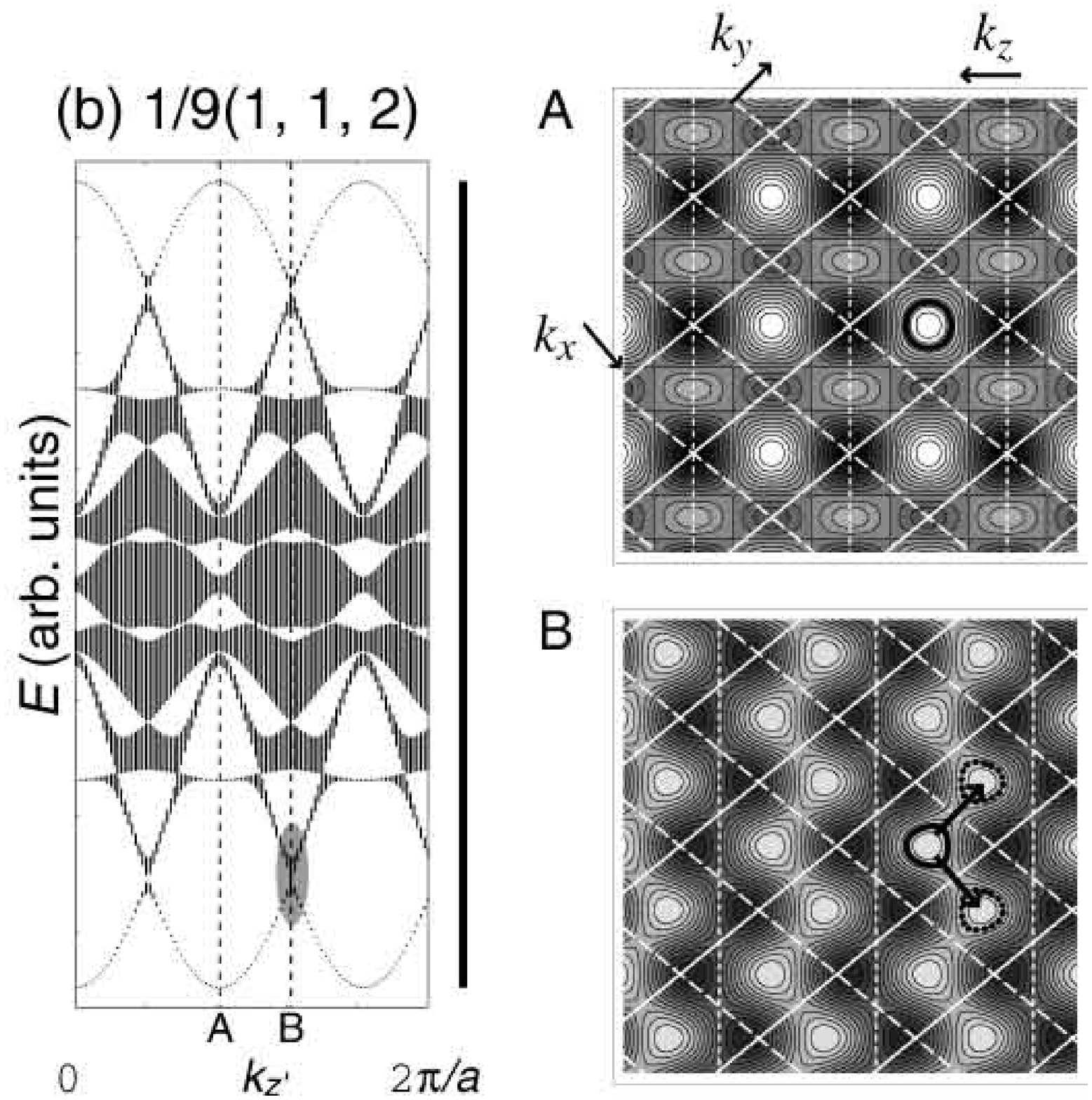}
\end{center}
\end{figure}
\begin{figure}
\begin{center}
  \leavevmode\epsfxsize=80mm \epsfbox{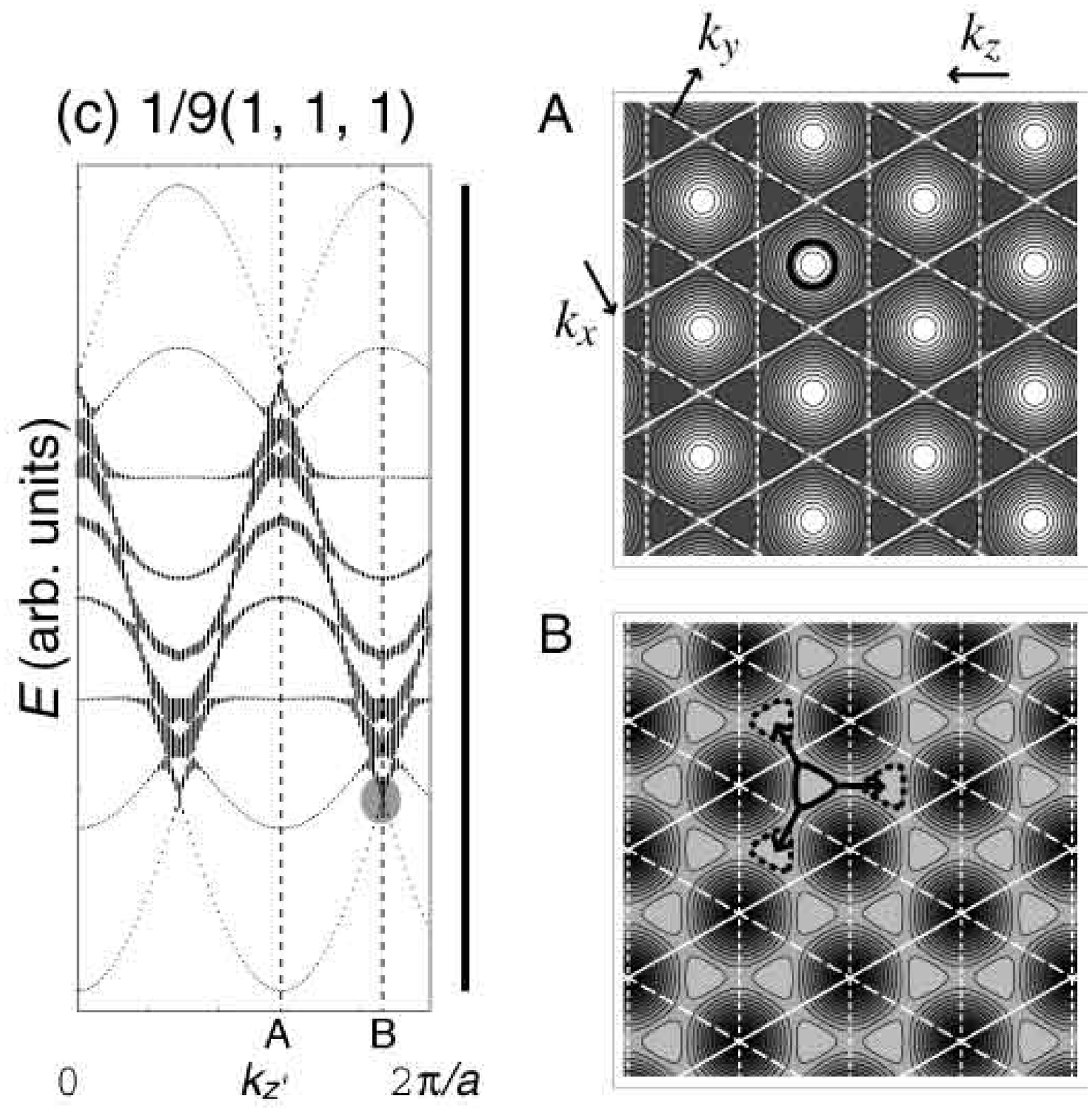}
\end{center}
\end{figure}
\begin{figure}
\begin{center}
  \leavevmode\epsfxsize=80mm \epsfbox{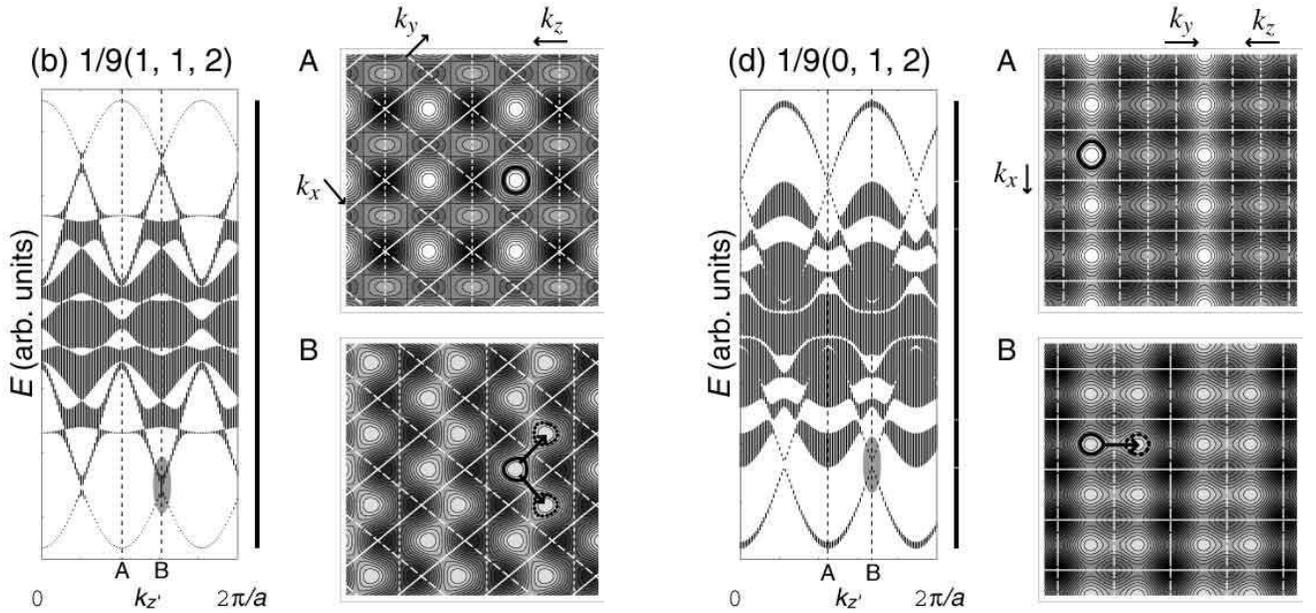}
\end{center}
\caption{
Energy spectra versus $k_{z'}$ for Eq.(\ref{HarpReduced2})
(left panels), where the final band structure is
indicated as bars at the right-hand edge of each spectrum.  
Right panels are grayscale plots of the fictitious potential
$\varepsilon(K_{x'},K_{y'},k_{z'})$ versus $(K_{x'},K_{y'})$ 
(with lighter areas corresponding to dips) for different values
of $k_{z'}$ (labeled as A, B, C in the left panels).  
Three sets of white lines represent the
intersections of the boundaries of the reciprocal unit cells
normal to $k_x,k_y,k_z$, respectively.  
We show three typical cases of $(\phi_x,\phi_y,\phi_z)=(1/9)(n_x, n_y, n_z)$: 
(a) $(n_x,n_y,n_z)=(1,2,3)$, (b) $(1,1,2)$, 
(c) $(1,1,1)$, and (d) $(0,1,2)$. The energy spectra are periodic in
$k_{z'}$ with period $2\pi/(an)$ with 
$n=\sqrt{n_x^2+n_y^2+n_z^2}$.  
Motion of the orbits with $k_{z'}$ for the
lowest subband (as well as for the second subband in (a)) 
are indicated, where the jumps to 
adjacent orbits are represented as arrows in the right panels
and the resulting energy gaps as shaded ovals in the left.  
In (a) the energy spectrum for a weak field ($(1/45)(1,2,3)$) 
are also shown for comparison.
}
\label{fig_band}
\end{figure}

\begin{figure}
\begin{center}
  \leavevmode\epsfxsize=80mm \epsfbox{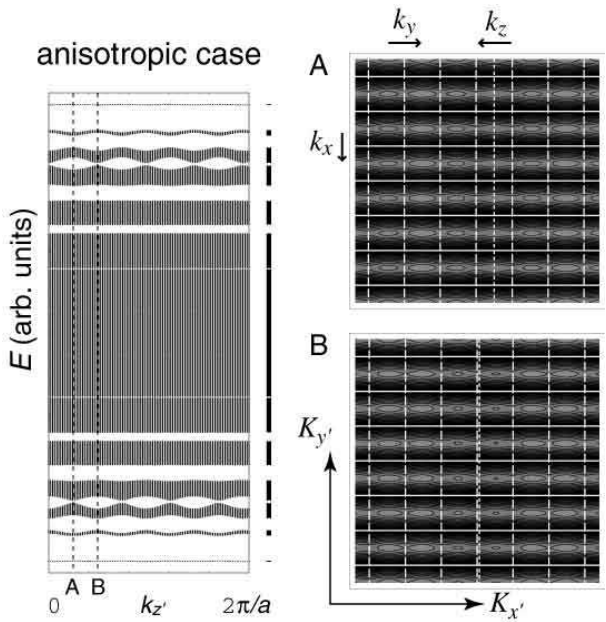}
\end{center}
\caption{
A plot similar to Fig.\ref{fig_band} 
for an anisotropic 3D system with $(t_x,t_y,t_z) = (1.0,0.1,0.1)$ 
for $(\phi_x,\phi_y,\phi_z)=(1/15)(0,1,4)$.  
The contours highlight the troughs.
}
\label{fig_aniso}
\end{figure}

In the previous paper\cite{Kosh} we have shown that an analogue of 
Hofstadter's butterfly arises (i) as a function of 
the {\it tilting angle} $\theta$ in magnetic field tilted in $yz$-plane 
(ii) in {\it anisotropic} ($t_x \gg t_y,t_z$) 3D lattices. 
If we apply the above argument to this case,
we see that the fictitious potential $\varepsilon$ is
dominated by the cosine band associated with the 
conductive direction, $x$.  As shown in Fig. \ref{fig_aniso} 
this gives rise to one-dimensional bound states in its
troughs and ridges with a small perturbation due to $t_y$ and $t_z$.  
So the system can be regarded as an array of independent chains 
with two periods, which is described by a Harper's equation
for the 2D Hofstadter's problem, 
and thus we can understand why the anisotropic case can have many 
gaps around the top and bottom of the band
as seen in Fig. \ref{fig_aniso},
while one component $\sigma_{yz}$ is fixed to 0
unlike the isotropic case.

\section{Dual Hall Conductivity}
We have just shown that two different systems 
related by the duality have similar energy spectra.
It would be then interesting to see
how the Hall conductivities are related between them 
for the corresponding energy gaps. 
For the 2D case, Thouless {\it et al}\cite{Thou} have shown the
relationship between the Hall integers
in the tight-binding and in the strong-field limits.
Here we first review this in the context of our duality
before moving on to the 3D case.
Diophantine's equation for an energy gap in the 2D tight-binding case  
(a 2D version of Eq.\ref{Dio3D}) is
\begin{equation}
 \nu_B = s + t \phi, 
\label{Dio-TB}
\end{equation}
where $s,t$ are integers, 
$\nu_B$ the tight-binding band filling,
and $\phi$ the number of fluxes penetrating the 2D unit cell.
The equation for the corresponding gap in the 2D strong-field case is
written with the same $s,t$ as
\begin{equation}
 \nu_L = s + t (1/\phi), 
\label{Dio-SF}
\end{equation}
where $\nu_L$ is the Landau-level filling factor, i.e., 
the fraction of the filled 
states in the lowest Landau level, 
and $\phi$ is replaced with $1/\phi$ according to the duality 
(Eq.\ref{duality}).  From the relation $\nu_L = \nu_B/\phi$, we can
translate Eq.(\ref{Dio-SF}) into
\begin{equation}
 \nu_B = s \phi + t.
\label{Dio-SF2}
\end{equation}
Now we can see that the two topological integers $s$ and $t$ 
are {\it interchanged} between the dual 
[(tight-binding (\ref{Dio-TB}) $\leftrightarrow$ 
strong-field (\ref{Dio-SF2})] cases.
The Widom-Str\v{e}da formula,\cite{Wido,Stre} 
\begin{equation}
\sigma_{xy} = -\frac{e^2}{h}\frac{\partial \nu_B}{\partial \phi}, 
\end{equation}
indeed 
dictates that the Hall conductivity is the gradient in $\nu_B$-$\phi$ diagram, 
which gives $-(e^2/h) t$ in the former case $-(e^2/h) s$ in the latter.

This results can be readily extend to the 3D case.
Namely, if we take an energy gap in the 3D system with 
$(\sigma_{yz},\sigma_{zx},\sigma_{xy})=-\frac{e^2}{ha}(m_x,m_y,m_z)$, 
the Hall conductivity in the corresponding 2D system becomes 
$-(e^2/h)s$, 
where $m_i$ and $s$ are related via Eq.(\ref{Dio3D}).

\section{Experimental feasibility}
Let us comment on the magnitude of the magnetic field required to
observe the energy gaps in 3D systems.  For that it is essential that
the coupling between the semiclassical orbits in the different wells 
exists.
For the isotropic crystals considered here, the area enclosed by the
semiclassical orbit in $\Vec{k}$-space should be of the order of the typical
well area, $2\pi / l^2 \sim (2\pi/a)^2$, which is simply $Ba^2 / \phi_0
\sim 1$.  The situation is the same as in the 2D Hofstadter problem, so
the required field for atomic lattice constants is huge ($B \sim 10^5$T
for $a=$2 \AA).  If we consider systems with larger unit cells, as
in solid fullerene or zeolites with $a=$10 \AA, the required $B$ 
is reduced to $10^3$T but still large.  
In the anisotropic case discussed in Ref.\cite{Kosh}, by 
contrast, the required is much less stringent.  
This is because the typical well area is $(2\pi/a)d$ with $d$ being the
typical valley width, so the condition relaxes to $2\pi / l^2 \sim
(2\pi/a)d$, which can be made as small as one wishes by increasing the
anisotropy (since $t_y, t_z \rightarrow 0$ leads to $d \rightarrow 0$),
although the scale of the energy gap shrinks.  So it is a trade-off
between large $B$ required with large energy gaps (isotropic case) 
and smaller $B$ suffices with small gaps (anisotropic).

Another point is that a real 3D sample has always surfaces.  
Halperin and the present authors have shown in general that 
the 3D integer QHE should accompany quantized {\it wrapping} 
current, whose intensity and direction are dictated by the 
quantum Hall integers\cite{KHA}. This should apply to the present 
case of isotropic QHE.

\section{Conclusion}
We have investigated the energy spectra in the isotropic 3D
lattice in magnetic fields applied in arbitrary directions, for which we have
shown that the energy gaps arise unless the magnetic field points to
high-symmetry directions.  We have also calculated the quantum Hall
integers.  In the latter part of the paper we have introduced a 
duality that relates $B \leftrightarrow 1/B$ (i.e., guiding 
center picture in strong $B$ and semiclassical picture in weak
$B$) for the present 3D problem, which gives the
graphic explanation of the quantization of the Hall conductivity
in 3D and the condition for the butterfly spectrum.  

M.K. would like to thank the JSPS for a financial support.

\end{multicols}
\end{document}